\documentclass[a4paper]{revtex4-1}
\usepackage{graphicx}
\usepackage{amsmath}
\usepackage{amssymb}
\usepackage{verbatim}

\newcommand{\boldnabla}{\mbox{\boldmath$\nabla$}}
\newcommand{\spin}{\mbox{\boldmath$\mathcal{S}$}}

\begin{document}

\title{Optical helicity of interfering waves}

\author{Robert P. Cameron, Stephen M. Barnett and Alison M. Yao }
\affiliation{Department of Physics, SUPA, University of Strathclyde, Glasgow G4 0NG, UK}
\email{robert.cameron@strath.ac.uk}

\begin{abstract}
Helicity is a property of light which is familiar from particle physics but less well-known in optics. In this paper we recall the explicit form taken by the helicity of light within classical electromagnetic theory and reflect upon some of its remarkable characteristics. The helicity of light is related to, but is distinct from, the spin of light. To emphasise this fact, we draw a simple analogy between the \textit{helicity} of light and electric {\it charge} and between the \textit{spin} of light and electric {\it current}. We illustrate this and other observations by examining various superpositions of plane waves explicitly.
\end{abstract}

\maketitle

\section{Introduction}
\label{introduction}

It was first suggested by Poynting \cite{Poynting09}, and later confirmed experimentally by Beth \cite{Beth36}, that a beam of light possesses a manifestly intrinsic `spin' angular momentum in the direction of propagation that is associated with the rotation of electric field vectors and magnetic field pseudovectors \cite{Barnett10,vanEnk94a,vanEnk94b}. This property of light, together with the `orbital' angular momentum that is associated with the spatial distribution of the electromagnetic field and its rotation \cite{Allen92}, has been the subject of much research interest in recent years \cite{Barnett10,OAM03,Yao11,Andrews13}.

Determining the fundamental description of the angular momentum of light in electromagnetic theory has proved surprisingly difficult. Regarding the manifestly intrinsic angular momentum of light in particular, clarity follows from an observation that is familiar from particle physics: the photon is massless and relativity suggests, therefore, that the photon only possesses one meaningful component of spin angular momentum, the component in the direction of propagation \cite{FeynmannIII06,vanEnk94a,vanEnk94b}. The value taken by this component is referred to as the photon's `\textit{helicity}'\cite{IntroParticle87}. The helicity of a photon in a circularly polarised plane wave mode is $\pm\hbar$, the plus or minus signs referring to left- or right-handed circular polarisation, respectively, in the optics convention \cite{ClassicalElectrodynamics99}. We suggest that it is the helicity, rather than the spin, of light that lies at the very heart of the description sought \cite{Cameron12a,Barnett12,Cameron12b}.

Helicity is a subtle property of light. Although it is closely related to the spin, and indeed the polarisation, of light, it is fundamentally distinct from both. It is somewhat difficult to convey the significance of such relationships and distinctions through an examination of the simplest optical field: a single plane wave. In this paper we instead consider more exotic optical fields composed of {\it multiple} plane waves. This allows us to explore the similarities and differences between the helicity and the energy of light and to illustrate an interesting analogy between the helicity of light and charge (both of which are signed quantities). Finally, it allows us to demonstrate clearly that, as claimed, the helicity of light is distinct from both the spin and polarisation of light.

\section{The optical helicity}
\label{Theopticalhelicity}

We work within the classical domain and consider freely-propagating light, adopting a system of units with $\epsilon_0=\mu_0=c=1$. The electric field, ${\bf E}$, and the magnetic field, ${\bf B}$, obey Maxwell's equations, which then take the form \cite{ClassicalElectrodynamics99}:
\sffamily
\begin{eqnarray}
\boldnabla\cdot{\bf  E}&=&0, \label{Gauss} \\
\boldnabla\cdot{\bf B}&=&0, \label{Gaussmagnetism} \\
\boldnabla\times{\bf E}&=&-\frac{\partial{\bf B}}{\partial t}, \label{FaradayLenz} \\
\boldnabla\times{\bf B}&=&\frac{\partial{\bf E}}{\partial t}. \label{AmpereMaxwell}
\end{eqnarray}
\normalfont

We begin by recalling some observations regarding the helicity of light that have been made elsewhere by ourselves and others  \cite{Cameron12a,Barnett12,Cameron12b,Candlin65,Calkin65,Zwanziger68,Deser76,Przanowski92,Przanowski94,Trueba96,Afanasiev96,Drummond99,Anco05,Drummond06,Fernandez-Corbaton12a,Fernandez-Corbaton12b,Coles12,Bliokh13}. The explicit expression for the total helicity of an optical field is
\begin{equation}
\mathcal{H}=\int\!\!\int\!\!\int \frac{1}{2}\left({\bf A}^\bot\cdot{\bf B}-{\bf C}^\bot\cdot{\bf E}\right) \, d^3{\bf r}, \label{theopticalhelicity}
\end{equation}
where ${\bf A}^\bot$ and ${\bf C}^\bot$ are, respectively, the transverse, gauge-invariant \cite{PhotonsAtomsWaves89} pieces of the familiar magnetic vector potential and an analogous electric pseudovector potential, due to Bateman \cite{Bateman15}, defined such that
\begin{equation}
{\bf E}=-\boldnabla\times{\bf C}^\bot=-\frac{\partial {\bf A}^\bot}{\partial t}, \ \ {\bf B}=\boldnabla\times{\bf A}^\bot=-\frac{\partial{\bf C}^\bot}{\partial t}. \label{potential}
\end{equation}
The optical helicity, $\mathcal{H}$, was introduced by Candlin \cite{Candlin65}, who referred to it as the `screw action'. It is a conserved, manifestly intrinsic quantity with the dimensions of an angular momentum, in spite of the fact that it is a time-even Lorentz pseudoscalar, possessing no sense of orientation in spacetime. The form of the operator, $\hat{\mathcal{H}}$, representing the optical helicity, $\mathcal{H}$, within the quantum domain was observed by Candlin \cite{Candlin65} and corresponds to the notion of helicity that is familiar from particle physics, describing a helicity of $\pm\hbar$ per photon in a circularly-polarised plane wave mode. The optical helicity, $\mathcal{H}$, is related to, but is ultimately distinct from, the `magnetic helicity' introduced by Woltjer \cite{Woltjer58}, which is utilized in the study of certain plasmas. Moreover, the form of the optical helicity, $\mathcal{H}$, is resembled by that of the `vortex helicity' or `fluid helicity' introduced by Moreau \cite{Moreau61}, which is utilized in the study of certain fluids\footnote{The use of the word `helicity' in this context is due to Moffatt \cite{Moffatt69} who proposed it by analogy with the concept of helicity familiar to him from particle physics.}. 

The helicity of light is, in fact, {\it locally} conserved, as expressed by the helicity continuity equation\footnote{We hope that no confusion arises here between the helicity density and Planck's constant: in the present paper, we use $h$ to denote the former and $\hbar$ to denote the reduced form of the latter.}:
\begin{equation}
\frac{\partial h}{\partial t}+\boldnabla\cdot{\bf s}=0, \label{helicitycontinuity}
\end{equation}
where
\begin{equation}
h=\frac{1}{2}\left( {\bf A}^\bot\cdot{\bf B}-{\bf C}^\bot\cdot{\bf E}\right) \label{helicitydensity}
\end{equation}
is the integrand seen in (\ref{theopticalhelicity}), which we interpret as being the \textit{helicity density} of an optical field, and we identify
\begin{equation}
{\bf s}=\frac{1}{2}\left({\bf E}\times{\bf A}^\bot+{\bf B}\times{\bf C}^\bot\right) \label{spinndensity}
\end{equation}
as being the {\it helicity flux density}. Remarkably, the helicity flux density, ${\bf s}$, is also the candidate for the {\it spin density} of an optical field that was put forward recently by one of us \cite{Barnett10}. Indeed, the volume integral of ${\bf s}$ is the total spin, $\spin$, of an optical field \cite{Barnett10,PhotonsAtomsWaves89,Bialynicki-Birula11,vanEnk94a,vanEnk94b}, the correct form of which is due to Darwin \cite{Darwin32,Bialynicki-Birula11} and to Humblet \cite{Humblet43,Marrucci11}. Although they are related, we emphasise that the optical helicity, $\mathcal{H}$, and the optical spin, $\spin$, are distinct: whereas the helicity of a photon in a circularly-polarised plane wave mode is $\pm\hbar$, its spin is $\pm\hbar\mathbf{k}/|\mathbf{k}|$. The latter result was observed by Lenstra and Mandel \cite{Lenstra82} and was investigated in greater detail by van Enk and Nienhuis \cite{vanEnk94a,vanEnk94b} who showed that the optical spin, $\spin$, is not a true angular momentum in that the operators representing its components within the quantum domain do not obey the usual angular momentum commutation relations. We have explored this point in detail elsewhere \cite{Cameron12a,Barnett10,Cameron12b}.

In playing its dual role, the helicity flux density or spin density, ${\bf s}$, reminds us of Poynting's vector, which also plays a dual role in that it is simultaneously the energy flux density and the linear momentum density of an optical field \cite{ClassicalElectrodynamics99}. Indeed, equation (\ref{helicitycontinuity}) is reminiscent of Poynting's theorem \cite{Poynting84}. We have considered the analogy between the fundamental descriptions in electromagnetic theory of the helicity of light and the energy of light in detail elsewhere \cite{Cameron12a,Barnett12}. 

The rather surprising appearance of {\it two} potentials in these discussions may be viewed as a reflection of something deeper: Maxwell's equations (\ref{Gauss}--\ref{AmpereMaxwell}) are highly symmetric. In particular, they place the electric field, ${\bf E}$, and the magnetic field, ${\bf B}$, on equal footing, retaining their form under a `duality rotation':
\begin{equation}
{\bf E }\rightarrow{\bf E}'= {\bf E}\cos\theta+{\bf B}\sin\theta, \ \ {\bf B}\rightarrow{\bf B}'= {\bf B}\cos\theta-{\bf E}\sin\theta, \label{dualityrotation}
\end{equation}
for any time-odd Lorentz pseudoscalar angle $\theta$, an observation due to Heaviside \cite{Heaviside92} and Larmor \cite{Larmor97}. It is this \textit{rotational symmetry} that is associated with the conservation of the helicity of light, a connection made by Calkin \cite{Calkin65} which ourselves and others have pursued elsewhere within the context of Noether's theorem \cite{Cameron12b,Bliokh13}. There is thus a sense in which the optical helicity, $\mathcal{H}$, embodies the principle of electric-magnetic democracy, a phrase coined by Berry \cite{Berry09}. Such associations seem reasonable given that a duality rotation (\ref{dualityrotation}) literally rotates the electric field vectors and magnetic field pseudovectors in a single plane wave about its direction of propagation \cite{Cameron12a,Barnett12,Cameron12b}. 

\section{The density of helicity: interference versus quasi-interference}
\label{analogy}

In section \ref{Theopticalhelicity}, we suggested an analogy between the helicity and energy of light. 
Comparing the helicity density, $h$, given by (\ref{helicitydensity}) and the energy density \cite{ClassicalElectrodynamics99},
\begin{equation}
w= \frac{1}{2}\left(\mathbf{E}\cdot\mathbf{E}+\mathbf{B}\cdot\mathbf{B}\right),
\label{energydensity}
\end{equation}
we see that they bear a remarkable resemblance \cite{Cameron12a}. Physically, however, they differ in many subtle respects.

In general, the energy density, $w$, is positive, although it can vanish at certain points in space at certain times. In contrast, the helicity density, $h$, can be positive, vanishing or \textit{negative}. 
The energy density, $w$, is sensitive to the phenomenon of {\it interference}, manifest in the presence of the dot products ${\bf E}\cdot{\bf E}$ and ${\bf B}\cdot{\bf B}$ seen in (\ref{energydensity}): interference is maximised when parallel fields are superposed but is absent when orthogonal fields are superposed. The helicity density, $h$, however, is \textit{not} inherently sensitive to interference but rather, we suggest by analogy, to a kind of `interference' between the electric and magnetic fields and the transverse pieces of their associated potentials, manifest in the presence of the dot products ${\bf A}^\bot\cdot{\bf B}$ and $-{\bf C}^\bot\cdot{\bf E}$ seen in (\ref{helicitydensity}). We refer to this phenomenon as {\it quasi-interference}.

\begin{figure}[h!]
\begin{center}
\includegraphics[scale=0.5]{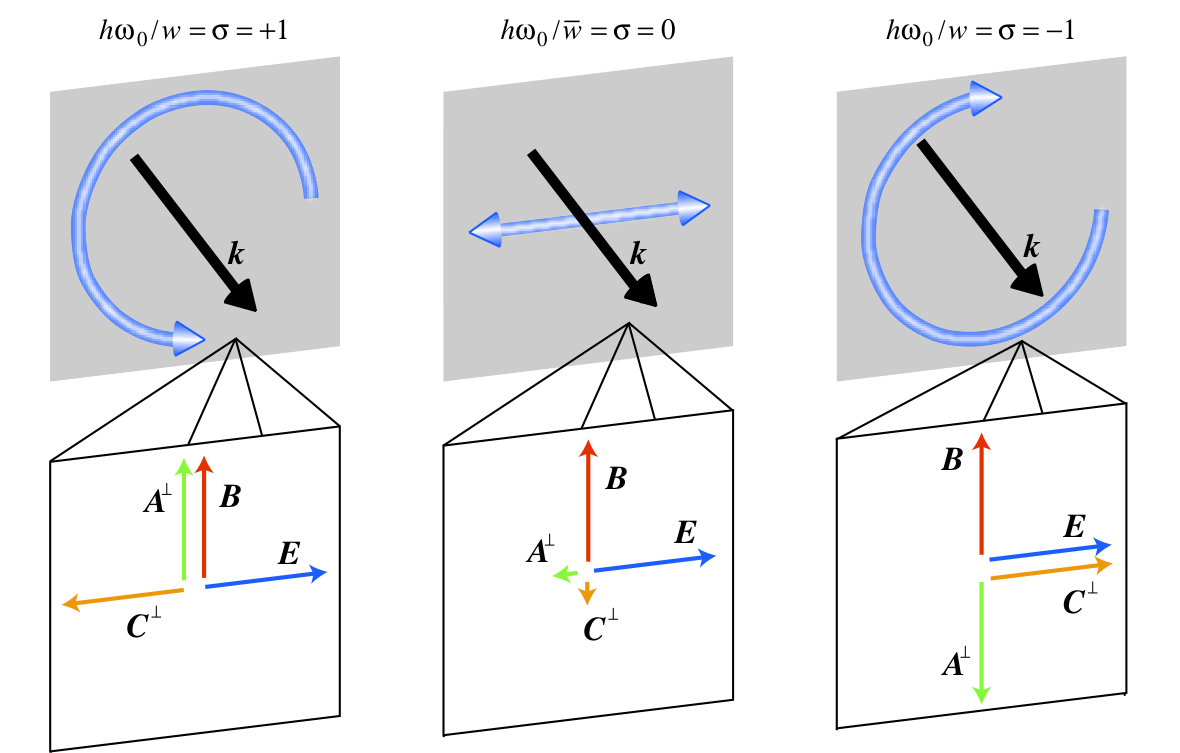}
\caption{\small Three different plane waves of equal amplitude and equal angular frequency, $\omega_0$. The polarisations of the plane waves are indicated by shaded (light blue online) arrows. The fields, ${\bf E}$ (blue online) and ${\bf B}$ (red online), and the transverse pieces, ${\bf C}^\bot$ (orange online) and ${\bf A}^\bot$ (green online), of their associated potentials, are depicted at some instant of time within the expanded views of the wavefronts. The value taken by the helicity density, $h$, given by (\ref{helicitydensity}) should be compared with the relative orientations of ${\bf E}$ and ${\bf C}^\bot$ and of ${\bf B}$ and ${\bf A}^\bot$ within each plane wave \cite{Barnett12}.}
\label{Figure1} 
\end{center} 
\end{figure}

We illustrate these ideas by considering an optical field composed of a superposition of two linearly-polarised plane waves, $1$ and $2$, of equal angular frequency, $\omega_0$. Initially, we suppose that the wavevectors, ${\bf k}_1$ and ${\bf k}_2$, of the plane waves are \textit{parallel}. Of course, the optical field itself may then be thought of as a single plane wave which can, in turn, be decomposed in various other ways however, these decompositions are not convenient for our present exposition. Now, suppose that the polarisations of the plane waves are parallel. The electric fields and the magnetic fields of the plane waves are also parallel and, therefore, the plane waves interfere. The electric and magnetic fields of each plane wave are, however, orthogonal to the transverse pieces of the associated potentials of the other plane wave and, therefore, the plane waves do not exhibit quasi-interference. The nature of the interference (constructive or destructive) is dictated by the relative phase of the plane waves which influences the amplitude of the (linearly-polarised) optical field. The energy density, $w$, of the optical field can be greater than, equal to or less than the sum of the energy densities that are attributable to the plane waves individually. The helicity density, $h$, of the optical field vanishes. If we suppose instead that the polarisations of the plane waves are orthogonal, the electric fields and the magnetic fields of the plane waves are also orthogonal and, therefore, the plane waves do not interfere. The electric and magnetic fields of each plane wave are, however, parallel to the transverse pieces of the associated potentials of the other plane wave and, therefore, the plane waves exhibit quasi-interference. The nature of the quasi-interference is also dictated by the relative phase of the plane waves which now influences the polarisation of the (elliptically-polarised in general) optical field. The energy density, $w$, of the optical field is simply the sum of the energy densities that are attributable to the plane waves individually. The helicity density, $h$, of the optical field however assumes a value `equivalent' to $\hbar \sigma$ per photon, where the polarisation parameter, $-1\le\sigma\le1$, takes its limiting values of $\pm 1$ for left- or right-handed circular polarisation, respectively \cite{Cameron12a, Barnett12}. If the optical field is of left-handed circular polarisation ($\sigma=+1$), it is found that ${\bf A}^\bot$ and ${\bf C}^\bot$ are in phase with, and are parallel to and anti-parallel to, ${\bf B}$ and ${\bf E}$, respectively, giving rise to a positive helicity density, $h$. Opposing relative orientations are found if the optical field is of right-handed circular polarisation ($\sigma=-1$), giving rise to a negative helicity density, $h$. If the optical field is linearly-polarised ($\sigma=0$), however, it is found that ${\bf A}^\bot$ and ${\bf C}^\bot$ are a quarter cycle out of phase with, and are orthogonal to, ${\bf B}$ and ${\bf E}$, respectively, giving rise to a vanishing helicity density, $h$. See figure \ref{Figure1}. Zambrini and Barnett have made observations that are closely related to those made here \cite{Zambrini07}. 

\begin{figure}[h!]
\begin{center}
\includegraphics[scale=0.6]{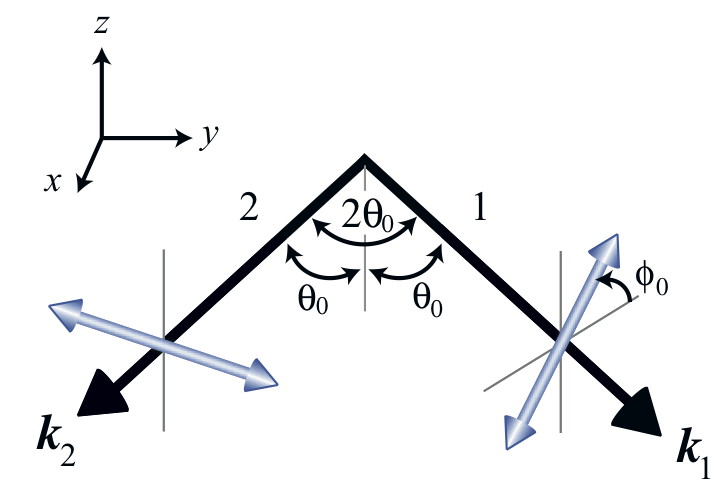}
\caption{\small Two plane waves, the wavevectors, ${\bf k}_1$ and ${\bf k}_2$, of which lie in the $x$-$y$-plane, separated by an angle $0<2\theta_0\ll1$. The superposition of these plane waves constitutes the optical field examined in the latter part of section \ref{analogy}. The angular separation of the wavevectors has been exaggerated for the sake of clarity.}
\label{Figure2} 
\end{center} 
\end{figure}

\begin{figure}[h!]
\begin{center}
\includegraphics[scale=0.65]{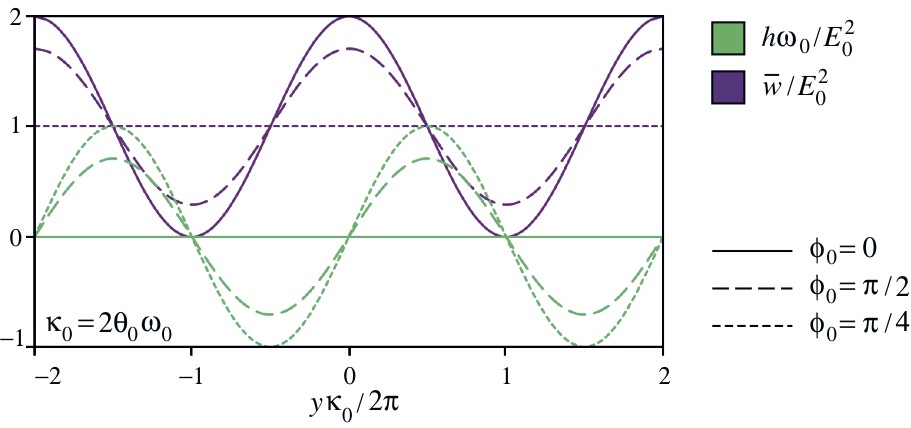}
\caption{\small From (\ref{resultsone}), plots of the helicity density, $h$, and the cycle-averaged energy density, $\overline{w}$, of the optical field depicted in figure \ref{Figure2}.}
\label{Figure3} 
\end{center} 
\end{figure}

Now let us consider what happens when the wavevectors, ${\bf k}_1$ and ${\bf k}_2$, of the plane waves lie within the $x$-$y$-plane, enclosing small angles $0<\theta_0\ll 1$ on either side of the $x$-axis such that ${\bf k}_1$ and ${\bf k}_2$ are themselves separated by an angle of $2\theta_0$ as depicted in figure \ref{Figure2}. 
In this particular case, we take the amplitudes of both plane waves to be equal to $E_0>0$, supposing that the polarisation of plane wave $2$ is confined to the $x$-$y$-plane whilst the polarisation of plane wave $1$ is at an angle $\phi_0$ to this plane. We take the wavevectors and the electric fields of the plane waves to be:
\begin{eqnarray}
{\bf k}_1 &=& \hat{{\bf x}} \cos \theta_0 + \hat{{\bf y}} \sin \theta_0 , \\
{\bf k}_2 &=& \hat{{\bf x}} \cos \theta_0 - \hat{{\bf y}} \sin \theta_0 ,  \\
{\bf E}_1 &=& \left ( \left (- \hat{{\bf x}} \sin \theta_0 + \hat{{\bf y}} \cos \theta_0 \right ) \cos \phi_0 +  \hat{{\bf z}} \sin \theta_0 \right ) E_0 \cos \left( {\bf k}_1 \cdot {\bf r} - \omega_0 t \right ), \\
{\bf E}_2 &=& \left (\hat{{\bf x}} \sin \theta_0 + \hat{{\bf y}} \cos \theta_0 \right ) E_0 \cos \left( {\bf k}_2 \cdot {\bf r} - \omega_0 t \right ),
\end{eqnarray}
where $\hat{{\bf x}}$, $\hat{{\bf y}}$ and $\hat{{\bf z}}$ are Cartesian unit vectors and the total electric field is of course given by ${\bf E}={\bf E}_1+{\bf E}_2$. To first order in $\theta_0$, the helicity density, $h$, and the \textit{cycle-averaged} energy density, $\overline{w}$, of the optical field are
\begin{eqnarray}
h &=& \frac{E_0^2}{\omega_0}\sin\phi_0 \sin\left(\kappa_0 y\right), \\
\overline{w} &=& E_0^2 \left(1+\cos\phi_0 \cos\left(\kappa_0 y\right)\right),  
\label{resultsone}
\end{eqnarray}
where $\kappa_0=2\theta_0 \omega_0$ is a wavenumber. Notice that the helicity density, $h$, is time independent. It can be shown that this is true of all strictly monochromatic optical fields: $\partial h / \partial t=0$. Due to the small angular separation, $2\theta_0$, of the plane waves, their relative phase oscillates as a function of $y$, with wavelength $2\pi/\kappa_0$. The helicity density (green online) and cycle-averaged energy density (purple online) are plotted in figure \ref{Figure3} for $\phi_0 = 0, \pi/4$ and $\pi/2$. For $\phi_0=0$ the polarisations of the plane waves are (essentially, as $\theta_0$ is small) parallel (see figure \ref{Figure2}) and the optical field is linearly polarised. The plane waves then interfere such that $\overline{w}$ oscillates between $2 E_0$ and $0$. That is, we have `bright' and `dark' fringes: a redistribution of energy within the optical field, attributable to constructive and destructive interference, respectively. In contrast, no quasi-interference occurs and $h$ vanishes (solid lines in figure \ref{Figure3}). An optical field possessing such characteristics can be found, for example, in the far-field of a Young's double slit diffraction pattern using linearly polarised light. If, instead, $\phi_0=\pi/2$, the polarisations of the plane waves are orthogonal and the polarisation parameter, $\sigma$, of the optical field oscillates between $\sigma=+1$ and $\sigma=-1$, reflecting the oscillation of the relative phase of the plane waves discussed above. As the plane waves do not interfere in this case, $\overline{w}$ assumes a value of $E_0^2$ at every point in space. The plane waves do, however, exhibit quasi-interference such that $h$ oscillates between $+E_0^2/\omega_0$ and $-E_0^2/\omega_0$ (see small dashed lines in figure \ref{Figure3}). That is, we have `helicity fringes': a redistribution of helicity about $h=0$ within the optical field, attributable to quasi-interference. Such optical fields have been utilised recently in optical trapping experiments \cite{Mohanty05,Cipparrone10} and have been referred to as polarisation gratings. Gradients in the polarisation of an optical field are also utilised, of course, in the laser cooling of certain atoms \cite{Ungar89,Dalibard89,Chu98,Cohen-Tannoudji98,Phillips98}. Although they are related, it should be borne in mind that the helicity and the polarisation of light are \textit{distinct}, a point to which we return in section \ref{helicityversuspolarisation}.

\section{The flow of helicity}
\label{Helicityversusspin}

In section \ref{Theopticalhelicity} we introduced the helicity flux density or spin density, ${\bf s}$, and made an analogy between the helicity continuity equation (\ref{helicitycontinuity}) and Poynting's theorem. Although we are considering freely-propagating light, assuming a strict absence of charge, we can also make a comparison with the familiar charge continuity equation \cite{ClassicalElectrodynamics99}:
\begin{equation}
\frac{\partial \rho}{\partial t}+\boldnabla\cdot {\bf j}=0,
\end{equation}
which relates the charge density, $\rho$, to the current density, ${\bf j}$. Charge, and indeed the charge density, $\rho$, can be positive or negative and a flow of positive charge in a given direction can yield the same current density, ${\bf j}$, as a suitable flow of negative charge in the opposite direction. Similarly, the helicity, and indeed the helicity density, $h$, of an optical field can be positive or negative and a flow of positive helicity in a given direction can yield the same helicity flux density, ${\bf s}$, as a suitable flow of negative helicity in the opposite direction. 
The analogy stops there, however, as there is no obvious physical significance to the volume integral of the current density, ${\bf j}$, whereas the volume integral of the helicity flux density or spin density, ${\bf s}$, is the optical spin, $\spin$, which is the total spin of an optical field \cite{Cameron12a,Barnett12,Cameron12b}, as discussed in section \ref{Theopticalhelicity}.

\begin{figure}[h!]
\begin{center}
\includegraphics[scale=0.5]{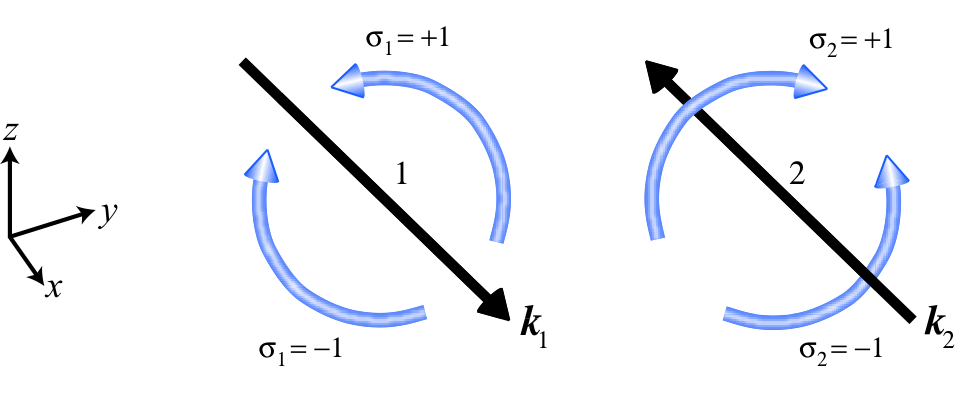}
\caption{\small Two circularly-polarised plane waves, the superposition of which constitutes the optical field examined in section \ref{Helicityversusspin}.}
\label{Figure4} 
\end{center} 
\end{figure}

To illustrate these ideas, let us consider the optical field depicted in figure \ref{Figure4}, which is composed of two circularly-polarised plane waves, $1$ and $2$, of equal amplitude, $E_0/\sqrt{2}>0$, and equal angular frequency, $\omega_0$, propagating in the $+x$- and $-x$-directions and possessing polarisation parameters $\sigma_1, \sigma_2\in\{-1,1\}$, respectively. The corresponding electric fields are
\begin{eqnarray}
{\bf E}_1 &=& \frac{E_0}{\sqrt{2}}  \left ( \hat{{\bf y}} \cos \left ( \omega_0 \left ( x - t \right ) \right ) - \hat{{\bf z}} \sigma_1 \sin \left ( \omega_0 \left ( x - t \right ) \right ) \right ),  \\
{\bf E}_2 &=& \frac{E_0}{\sqrt{2}}  \left ( - \hat{{\bf y}} \cos \left ( \omega_0 \left ( x - t \right ) \right ) + \hat{{\bf z}} \sigma_2 \sin \left ( \omega_0 \left ( x + t \right ) \right ) \right ),
\end{eqnarray}
and the total electric field is of course ${\bf E} = {\bf E}_1 + {\bf E}_2$. The helicity density, $h$, and the helicity flux density or spin density, ${\bf s}$, of the optical field are:
\begin{eqnarray}
h &=& \frac{E_0^2}{2 \omega_0} \left(\sigma_1+\sigma_2 \right), \\
{\bf s} &=& \frac{E_0^2}{2 \omega_0} \left(\sigma_1-\sigma_2 \right)\hat{{\bf x}}.
\label{spindensityintwowaves}
\end{eqnarray}
Notice that, like the helicity density, $h$, the helicity flux density or spin density, ${\bf s}$, is time independent.  It can be shown that this is true of all strictly monochromatic optical fields: $\partial {\bf s} / \partial t=0$.

If both plane waves possess the \textit{same} sense of circular polarisation ($\sigma_1=\sigma_2=\pm1$), there is a non-vanishing helicity density, $h=\pm E_0^2 /\omega_0$, but a vanishing helicity flux density or spin density, ${\bf s}=0$. In contrast, if the plane waves possess \textit{opposite} circular polarisations ($\sigma_1 =-\sigma_2 = \pm1$), there is a vanishing helicity density, $h=0$, but a non-vanishing helicity flux density or spin density, ${\bf s}=\pm  E_0  \hat{\boldsymbol{x}}/ \omega_0$.

Returning to the analogy made above between the helicity of light and charge, we can liken the first case ($\sigma_1=\sigma_2=\pm1$) to a combination of two counterpropagating flows of charge of the \textit{same} sign, giving rise to a net charge (c.f. $h\ne0$) but no net current (c.f. ${\bf s =0}$). In contrast, we can liken the second case ($\sigma_1 =-\sigma_2 = \pm1$) to a combination of two counterpropagating flows of charge of \textit{opposite} signs, yielding overall neutrality (c.f. $h=0$) whilst giving rise to a net current (c.f. ${\bf s}\ne0$). Counterpropagating circularly-polarised beams of light possessing the \textit{same} handedness are utilised in the process of cooling certain atoms \cite{Ungar89,Dalibard89,Chu98,Cohen-Tannoudji98,Phillips98} whilst counterpropagating circularly-polarised beams of light possessing \textit{opposite} handedness (but slightly different amplitudes) have been utilised recently in a fluorescence detected circular dichroism experiment \cite{Tang10,Tang11,Smart11}, which yielded an enhancement of a certain measure of dissymmetry over that which can be observed utilising a single traveling beam of circularly-polarised light.

Evidently, it is possible to produce light that possesses a non-vanishing helicity but a vanishing helicity flux and, in particular, a vanishing spin (and vice-versa). This is, we suggest, a clear demonstration that the helicity of light and the spin of light are indeed distinct, in spite of the intimate relationship between them that is embodied in the helicity continuity equation (\ref{helicitycontinuity}).

\section{A subtle distinction}
\label{helicityversuspolarisation}

In this section we explore the subtle distinction between the polarisation of light and the helicity of light. Polarisation is a concept which is invoked to describe the manner in which the {\it electric} field vectors, in particular, evolve in time within an optical field. In contrast, the helicity of light possesses the dimensions of an angular momentum and may be thought of as flowing continuously within an optical field, in accordance with the helicity continuity equation (\ref{helicitycontinuity}). Nevertheless, the polarisation and helicity density of a single plane wave are closely related \cite{Cameron12a,Barnett12}, as discussed in section \ref{analogy}. In particular, we can associate the presence of rotating {\it electric} field vectors (circular polarisation) directly with a non-vanishing helicity density. For more exotic optical fields, however, such associations are not necessarily appropriate. 

\begin{figure}[h!]
\begin{center}
\includegraphics[scale=0.6]{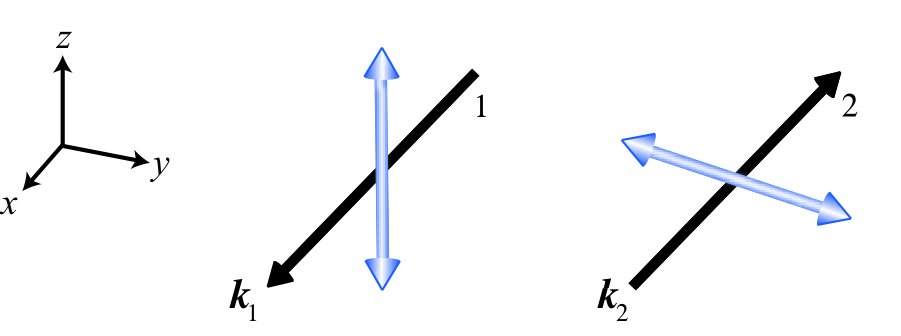}
\caption{\small Two plane waves, the superposition of which constitutes the optical field examined in section \ref{helicityversuspolarisation} and depicted in figure \ref{Figure6}.}
\label{Figure5} 
\end{center} 
\end{figure}

\begin{figure}[h!]
\begin{center}
\includegraphics[scale=0.575]{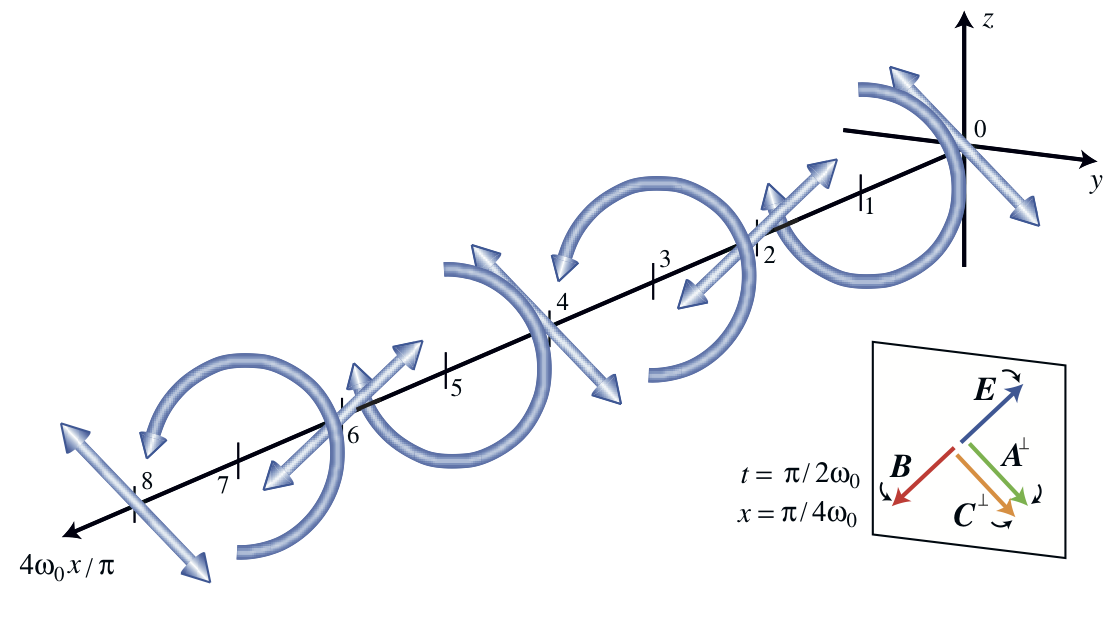}
\caption{\small The optical field resulting from the superposition depicted in figure \ref{Figure5}. The electric field vectors and the magnetic field pseudovectors within the optical field rotate with {\it opposite} senses.}
\label{Figure6} 
\end{center} 
\end{figure}

We can demonstrate this fact through an examination of the optical field depicted in figure \ref{Figure5}, which is composed of two plane waves, $1$ and $2$, of equal amplitude, $E_0>0$, and equal angular frequency, $\omega_0$, that are linearly-polarised parallel to the $z$- and $y$-axes and propagate in the $+x$- and $-x$-directions, respectively. Their electric fields are 
\begin{eqnarray}
{\bf E}_1 &=& \hat{{\bf z}} E_0 \cos \left ( \omega_0 \left ( x - t \right ) \right ), \\
{\bf E}_2 &=& -\hat{{\bf y}} E_0 \cos \left ( \omega_0 \left ( x + t \right ) \right ).
\end{eqnarray}
The total electric field, ${\bf E}={\bf E}_1+{\bf E}_2$, and indeed the total magnetic field, ${\bf B}$, can be expressed in the revealing forms:
\begin{eqnarray}
{\bf E}&=& \left(-\hat{{\bf y}}+\hat{{\bf z}}\right) E_0\cos\left(\omega_0 x\right)\cos\left(\omega_0 t\right) +\left(\hat{{\bf y}}+\hat{{\bf z}}\right)E_0\sin\left(\omega_0 x\right)\sin\left(\omega_0 t\right), \label{totalelectricfield} \\
{\bf B}&=&\left(-\hat{{\bf y}}+\hat{{\bf z}}\right) E_0 \cos\left(\omega_0 x\right)\cos\left(\omega_0 t\right) -\left(\hat{{\bf y}}+\hat{{\bf z}}\right) E_0 \sin\left(\omega_0 x\right)\sin\left(\omega_0 t\right) \label{linlinfield} .
\end{eqnarray}
Notice that the optical field is in the `lin $\bot$ lin' configuration as shown in figure \ref{Figure6}. Such fields may be utilised, for example, in the laser cooling of certain atoms, due to their inherent `polarisation gradients' \cite{Ungar89,Dalibard89,Chu98,Cohen-Tannoudji98,Phillips98}: at $x=x_N=N\pi  / 2 \omega_0$ and $x=x_M=\left(2 M+1\right)\pi/4 \omega_0$, the electric field vectors oscillate within the $y$-$z$-plane in linear and circular manners, 
respectively, where $N=0,\pm1,\pm2,\dots$ and $M=0,\pm1,\pm2,\dots$ are integers. The magnetic field pseudovectors also oscillate within the $y$-$z$-plane in linear and circular manners at $x=x_N$ and $x=x_M$,  respectively. However, the \textit{sense} of rotational motion that they exhibit is \textit{opposite} to that exhibited by the electric field vectors, due to the opposite signs of the second terms seen in (\ref{totalelectricfield}) and (\ref{linlinfield}). The optical field is depicted in figure \ref{Figure6}.

In this case we find that the helicity density, $h$, and the helicity flux density or spin density, ${\bf s}$, of the optical field both vanish:
\begin{equation}
h=0, \ \ {\bf s}=0.\label{helicitydensity2}
\end{equation}
This result is perhaps unsurprising given the {\it opposing} rotational motions exhibited by the electric field vectors and the magnetic field pseudovectors within the optical field: as discussed in section \ref{Theopticalhelicity}, the optical helicity, $\mathcal{H}$, itself lacks any sense of orientation in spacetime and embodies the principle of electric-magnetic {\it democracy}. Evidently, the mere existence of rotating electric field vectors (circular polarisation) does not in itself imply the existence of a non-vanishing helicity density, $h$, and helicity flux density or spin density, ${\bf s}$, within an optical field in general\footnote{Although the electric and magnetic fields are treated in an equal manner by Maxwell's equations (\ref{Gauss}--\ref{AmpereMaxwell}), they are, of course, distinct entities. In spite of the opposing senses of rotational motion exhibited by their vectors and pseudovectors, respectively, it would certainly {\it not} be fair to say that the optical field under examination possesses no rotational motion whatsoever: the electric and magnetic fields do not `cancel each other out'. Indeed, further investigation reveals the presence of non-vanishing components of spin flux density or ij-infra-zilches densities, $n_{ij}$ \cite{Cameron12a}.}.

\section{Discussion}

We have explored some of the relationships and distinctions between the helicity of light and other properties of light that may be more familiar, namely the energy, spin and polarisation of light. We have illustrated our observations by examining various superpositions of plane waves explicitly.

It should be noted that the optical helicity, $\mathcal{H}$, is {\it not} the only quantity in electromagnetic theory that describes the manifestly intrinsic angular momentum associated with the rotation of electric field vectors and magnetic field pseudovectors. In addition, we recognise the optical spin, $\spin$, and other quantities besides \cite{Cameron12a,Barnett12,Cameron12b}. In consideration of our recent investigations, it now appears that the description extends indefinitely. It seems clear, however, that the optical helicity, $\mathcal{H}$, lies at the very heart of this description, as suggested in section \ref{introduction}.

It is natural to ask, of course, whether the demonstrated ability to ``sculpt'' the helicity of an optical field is of any practical use, a subtle point given that helicity is {\it not} a conserved property of light in the presence of charge \cite{Barnett12}. We shall return to this interesting question in future publications.

\vskip1cm
\noindent {\bf Acknowledgements}
\vskip0.5cm
\noindent This work was supported by the Carnegie Trust for the Universities of Scotland, the U.K. Engineering and Physical Sciences Research Council (E.P.S.R.C) and the Leverhulme Trust. 



\begin{thebibliography}{38} 

\bibitem{Poynting09} 
J. H. Poynting, Proc. R. Soc. London Ser. A {\bfseries 82} 560--7 (1909) 
 
\bibitem{Beth36} 
R. A. Beth, Phys. Rev. {\bfseries 50} 115--25 (1936) 

\bibitem{Barnett10} 
S. M. Barnett, J. Mod. Opt. {\bfseries 57} 1339--43 (2010) 

\bibitem{vanEnk94a} 
S. J. van Enk and G. Nienhuis, Europhys. Lett. \textbf{25} 497--501(1994)

\bibitem{vanEnk94b} 
S. J. van Enk and G. Nienhuis,  J. Mod. Opt. \textbf{41} 963--77 (1994)

\bibitem{Allen92} 
L. Allen, M. W. Beijersbergen, R. J. C. Spreeuw and J. P. Woerdman, Phys. Rev. A {\bfseries 45} 8185--9 (1992) 

\bibitem{OAM03} 
L. Allen, S. M. Barnett and M. J. Padgett, {\it Optical Angular Momentum}, Institute of Physics Publishing: Bristol (2003) 

\bibitem{Yao11} 
A. M. Yao and M. J. Padgett, Adv. Opt. Phot. {\bfseries 2} 161--204 (2010)

\bibitem{Andrews13} 
D. L. Andrews and M. Babiker, {The Angular Momentum of Light}, Cambridge University Press: Cambridge (2013)

\bibitem{FeynmannIII06} 
R. P. Feynman, R. B. Leighton and M. Sands, {\it The Feynman Lectures on Physics: The Definitive Edition, Volume III}, Addison-Wesley: Westford, MA (2006) 

\bibitem{IntroParticle87} 
D. Griffiths, {\it Introduction to elementary particles}, Wiley: Chichester (1987) 

\bibitem{ClassicalElectrodynamics99} 
J. D. Jackson, {\it Classical Electrodynamics}, Wiley: New York (1999) 

\bibitem{Cameron12a} 
R. P. Cameron, S. M. Barnett and A. M. Yao, New J. Phys. {\bfseries 14} 053050 (2012) 

\bibitem{Barnett12} 
S. M. Barnett, R. P. Cameron and A. M. Yao, Phys. Rev. A {\bfseries 86} 013845 (2012) 

\bibitem{Cameron12b} 
R. P. Cameron and S. M. Barnett, New J. Phys. {\bfseries 14} 123019 (2012) 

\bibitem{Candlin65} 
D. J. Candlin, Il Nuovo Cimento {\bfseries 37} 1390--5 (1965) 

\bibitem{Calkin65} 
M. G. Calkin, Am. J. Phys {\bfseries 33} 958--60 (1965) 

\bibitem{Zwanziger68} 
D. Zwanziger, Phys. Rev. {\bfseries 176} 1489--95 (1968)

\bibitem{Deser76} 
S. Deser and C. Teitelboim, Phys. Rev. D {\bfseries 13} 1592--7 (1976)

\bibitem{Przanowski92} 
M. Przanowski and A. Macio\l{}ek-Nied$\dot{\mbox{z}}$wiecki, J. Math. Phys. {\bfseries 33} 3978--82 (1992)

\bibitem{Przanowski94} 
M. Przanowski, B. Rajca and J. Tosiek, Acta Phys. Pol. B {\bfseries 25} 1065--77 (1994) 

\bibitem{Trueba96} 
J. L. Trueba and A. F. Ra\~{n}ada, Eur. J. Phys. {\bfseries 17} 141--4 (1996)

\bibitem{Afanasiev96}  
G. N. Afanasiev and Y. P. Stepanovsky, Il Nuovo Cimento {\bfseries 109 A} 271--9 (1996)

\bibitem{Drummond99} 
P. D. Drummond, Phys. Rev. A, {\bfseries 60} R3331--4 (1999)

\bibitem{Anco05} 
S. C. Anco and D. The, Acta Appl. Math. {\bfseries 89} 1--52 (2005)

\bibitem{Drummond06} 
P. D. Drummond, J. Phys. B: At. Mol. Opt Phys. {\bfseries 39} S573–98 (2006)

\bibitem{Fernandez-Corbaton12a} 
I. Fernandez-Corbaton, X. Zambrana-Puyalto, N. Tischler N, {\it et al.}, arXiv:1206.0868v1 (2012)

\bibitem{Fernandez-Corbaton12b} 
I. Fernandez-Corbaton, X. Zambrana-Puyalto and G. Molina-Terriza, Phys. Rev. A {\bfseries 86} 042103 (2012)

\bibitem{Coles12} 
M. M. Coles and D. L. Andrews, Opt. Lett. {\bfseries 38} 869--71 (2013)

\bibitem{Bliokh13}
K. Y. Bliokh, A. Y. Bekshaev and F. Nori, New J. Phys. {\bfseries 15} 033026 (2013)

\bibitem{PhotonsAtomsWaves89} 
C. Cohen-Tannoudji, J. Dupont-Roc and G. Grynberg, {\it Photons and Atoms: Introduction to Quantum Electrodynamics}, Wiley: New-York (1989) 

\bibitem{Bateman15} 
H. Bateman, {\it The Mathematical Analysis of Electrical and Optical Wave-Motion on the Basis of Maxwell's Equations}, Cambridge University Press: Cambridge (1915) 

\bibitem{Woltjer58}  
L. Woltjer, Proc. Nat. Acad. Sci. {\bfseries 44} 489--91 (1958) 

\bibitem{Moreau61} 
J. J. Moreau, C. R. Acad. Sci. Paris {\bfseries 252} 2810--2 (1961) 

\bibitem{Moffatt69} 
H. K. Moffatt, J. Fluid Mech. {\bfseries 35} 117--29 (1969) 

\bibitem{Darwin32} 
C. G. Darwin, Proc. R. Soc. Lond. A {\bfseries 136} 36--52 (1932) 

\bibitem{Bialynicki-Birula11}
I. Bia\l{}ynicki-Birula and Z. Bia\l{}ynicka-Birula, J. Opt {\bfseries 13} 064014 (2011)   

\bibitem{Humblet43} 
J. Humblet, Physica X, {\bfseries 7} 585--603 (1943) 

\bibitem{Marrucci11} 
L. Marrucci, E. Karimi, S. Slussarenk, \textit{et al}., J. Opt {\bfseries 13} 064001 (2011) 

\bibitem{Lenstra82}
D. Lenstra and L. Mandel, Phys. Rev. A {\bfseries 25} 3428--37 (1982) 

\bibitem{Poynting84} 
J. H. Poynting, Phil. Trans. R. Soc. {\bfseries 175} 334--61 (1884) 

\bibitem{Heaviside92} 
O. Heaviside, Phil. Trans. R. Soc. A {\bfseries 183} 423--80 (1892) 

\bibitem{Larmor97} 
J. Larmor, Phil. Trans. R. Soc. A {\bfseries 190} 205--300 (1897) 

\bibitem{Berry09} 
M. V. Berry, J. Opt. A: Pure Appl. Opt. {\bfseries 11} 094001 (2010) 

\bibitem{Zambrini07} 
R. Zambrini and S. M. Barnett, Opt. Expr. {\bfseries 15} 15214--27 (2007) 

\bibitem{Mohanty05} 
S. K. Mohanty, K. D. Rhao and P. K. Gupta, Appl. Phys. B {\bfseries 80} 631--4 (2005) 

\bibitem{Cipparrone10} 
G. Cipparrone, I. Ricardez-Vargas, P. Pagliusi, \textit{et al}., Opt. Soc. Am. {\bfseries 18} 6008--13 (2010) 

\bibitem{Ungar89} 
P. J. Ungar, D. S. Weiss, E. Riss and S. Chu, J. Opt. Soc. Am. B {\bfseries 6} (1989)

\bibitem{Dalibard89} 
J. Dalibard and C. Cohen-Tannoudji, Opt. Soc. Am. {\bfseries 6} 2023--45 (1989) 

\bibitem{Chu98} 
S. Chu, Rev. Mod. Phys. {\bfseries 70}, 685--706 (1998)

\bibitem{Cohen-Tannoudji98} 
C. N. Cohen-Tannoudji, Rev. Mod. Phys. {\bfseries 70} 707--19 (1998)

\bibitem{Phillips98} 
W. D. Phillips, Rev. Mod. Phys. {\bfseries 70} 721--41 (1998)

\bibitem{Tang10} 
Y. Tang and A. E. Cohen, Phys. Rev. Lett. {\bfseries 104} 163901 (2010) 

\bibitem{Tang11} 
Y. Tang and A. E. Cohen, Science {\bfseries 332} 333--6 (2011) 

\bibitem{Smart11} 
A. G. Smart, Phys. Today, {\bfseries 64} 16--7 (2011) 

\end{thebibliography}
\end{document}